\documentstyle[preprint,aps]{revtex}

\begin{document}

\draft

\title{Features  of  first  passage  time  density function for
coherent stochastic  resonance  in  the  case  of  two  absorbing
boundaries}

\author{Asish K. Dhara and S. R. Banerjee}

\address{Variable  Energy  Cyclotron  Centre,  1/AF Bidhan Nagar,
Calcutta-700064, India}

%\tighten
\date{\today}
\maketitle

\begin{abstract}
Coherent  stochastic  resonance  is  explained  in terms of first
passage time density functions. Scaling  relation  between  first
passage time density functions at resonance for different lengths
of  the  medium  is  obtained.  A  formula for first passage time
density  function  at  resonance  is  derived  in  terms  of  two
universal  functions, which demonstrates the universal feature of
coherent stochastic resonance. A  simple  approximate  expression
for  first passage time density function at resonance is proposed
which is shown  to  explain  the  behavior  at  resonance  fairly
accurately.
\end{abstract}

KEY  WORDS: Synchronization, coherent stochastic resonance, first
passage time density functions


\section{introduction}

There  has  been a large deal of interest in the understanding of
mechanism of interplay between random noise and  a  deterministic
periodic signal. It has been found that they act cooperatively in
some   nonlinear  systems  \cite{bul1,gri}.  Nature,  presumably,
exploits this fact to its advantage  to  tune  in  to  a  desired
signal  \cite{dou,bul2}.  Even  for  linear,  spatially  extended
stochastic system the  cooperative  response  between  noise  and
deterministic  periodic force has been demonstrated very recently
\cite{fle,por,nic,dha}.

The  first  passage  time  is  a  useful  tool to investigate the
diffusive transport property in a medium.  The  theory  of  first
passage  time  has  been  worked  out  in  great  detail for both
infinite  medium  and   explicitly   time-independent   diffusive
processes    \cite{wei,gar,van}.    However,    for    explicitly
time-dependent processes and in finite medium an analytic  closed
form  expressions  are  not  available. In this respect also this
problem attracts much attention to the scientific communities.

The process that we are concerned is an overdamped system, having
no  intrinsic  frequency, driven by a periodic force and embedded
in a noisy environment which is taken to be  Gaussian  white  for
simplicity.  The  motion  is  constrained  between two traps. The
first analysis of this process has been done for a random walk on
a lattice numerically and in a continuous medium with a  periodic
signal   of  small  amplitude  perturbatively  \cite{fle}.  Their
results indicate that the oscillating field can create a form  of
coherent  motion  capable of reducing the first passage time by a
significant amount. In order to investigate the reason  for  this
cooperative  behavior  of random noise and deterministic periodic
signal, this problem has been formulated in much simpler terms by
approximating the sinusoidal periodic  signal  by  the  telegraph
signal  and  later  it  has  been  concluded  \cite{por} that the
telegraph  signal  cannot  produce  any   cooperative   behavior.
Subsequently, it has been shown \cite{dha} that if we approximate
the  sinusoidal signal by a multi-step periodic signal [explained
below] one would  recover  the  coherent  motion.  These  authors
\cite{dha} have also demonstrated why single-step telegraph signal
cannot  produce any coherent motion as conjectured by the earlier
worker \cite{por} and multi-step telegraph signal does.

While  some  authors  \cite{nic}  demonstrated the non-monotonous
behavior of the mobility as a function of  forcing  frequency  by
calculating  the  net  mass  leaving at one of the boundaries for
weak force, others have demonstrated the non-monotonous  behavior
of  the  mean  first  passage  time  (MFPT)  for  small amplitude
\cite{fle} and for arbitrary amplitude \cite{dha} in a continuous
medium. These authors \cite{dha} have also demonstrated that  the
variance  of the first passage time density function (FPTDF) also
attains a minimum at the same frequency for which the MFPT  shows
a  minimum  when  they  are  plotted  as  a  function  of forcing
frequency. This happened for any length of the  medium.  Such  an
enhancement  of  the  mobility at some definite forcing frequency
depending on the  length  of  the  medium  clearly  exhibits  the
maximum  cooperation  between  noise  and  periodic signal and is
known in the literature as coherent stochastic resonance (CSR).

All  the  previous analyses \cite{fle,por,dha} concentrate mainly
on the first order  moment  (MFPT)  of  the  first  passage  time
density  function (FPTDF) except in \cite{dha}, where it has been
shown that the FPTDF at resonance at  different  lengths  of  the
medium  overlap  over  each  other  when  they  are  plotted as a
function of the  cycle  of  the  periodic  signal.  This  clearly
indicates that the resonance feature is, in some sense, universal
in  nature. To investigate this feature further, we in this paper
concentrate on the full profile of FPTDF and study the effect  of
synchronization  between  noise and deterministic periodic signal
in terms of FPTDF as we change the frequency. It is shown that  a
single  dominant  peak  of  FPTDF  occurs  at resonance where the
synchronization is maximum. This feature is quite distinct and is
not observed when the system is away from the resonance. Further,
a scaling relation is obtained between FPTDFs  at  resonance  for
different lengths so that one could map FPTDF at arbitrary length
on  a  single  FPTDF at our choice by such scaling relation. This
feature together with the cycle variable of the  periodic  signal
as  argument  of FPTDF conclude the universal feature of FPTDF at
resonance.

The  paper  is  organised  as  follows.  In  Sec.II,  we give the
formulation of the problem. The results of the calculations based
on the derived formulae are presented in Sec.III. Apart from  the
explanation  of  synchronization  in  terms of FPTDF, the scaling
behavior of FPTDF is also presented in this section. The FPTDF at
resonance is expressed in terms of two universal functions, which
clearly isolates its dependence on  the  length  of  the  medium.
Further,  a  simple  expression  of  FPTDF  at  resonance is also
obtained in the same section which is shown to be fairly accurate
in  explaining  the  behavior  at  resonance.  Finally,   a   few
concluding remarks have been added in Sec.IV.

\section{Formulation of the problem}

The  system  that  we  consider  is  described by a Fokker Planck
equation

\begin{equation}
\label{eq.1}
\frac{\partial p(x,t)}{\partial t} = - A sin\Omega t\frac{\partial p(x,t)}
{\partial x} + D\frac{\partial^2 p(x,t)}{\partial x^2},
\end{equation}

where  $A$  and  $\Omega$  are the amplitude and frequency of the
sinusoidal signal, and $D$ is the strength of the Gaussian  white
noise. The motion is confined between two absorbing boundaries at
$x=0$  and  $x=L$  ;  i.e.,  Eq.(\ref{eq.1}) is supplemented with
absorbing boundary conditions $p(0,t)=p(L,t)=0$. In terms of  the
dimensionless variables

\begin{equation}
\label{eq.2}
\xi=(A/D)x , \theta = (A^2/D)t , \omega = \Omega/(A^2/D) ,
\end{equation}
the Eq.(\ref{eq.1}) is conveniently written as

\begin{equation}
\label{eq.3}
\frac{\partial p(\xi,\theta)}{\partial\theta} = - sin\omega\theta
\frac{\partial p(\xi,\theta)}{\partial\xi} +\frac{\partial^2p(\xi,\theta)}
{\partial \xi^2}.
\end{equation}
The boundary conditions read in terms of the new variables (2) as
$p(0,\theta)=p(\Lambda,\theta)=0$,  where  $\Lambda = (A/D)L$. In
the following we calculate all the physical quantities  in  terms
of these new variables and if required, one may translate all the
interpretations   in   terms   of  the  usual  variables  by  the
transformation equations Eq.(\ref{eq.2}).

No analytic solution exists for Eq.(\ref{eq.3}) with the boundary
conditions   as   mentioned.   We  thus  introduce  a  scheme  to
approximate the  force  $\sin  \omega  \theta$  as  a  multi-step
periodic  signal \cite{dha}.  This  scheme  is  in contrast to the procedure
adopted in \cite{fle}, where they discretise the  Eq.(\ref{eq.3})
using  finite  difference  method  and  simulate the problem on a
lattice  of  space  and  time.  The  construction  of  multi-step
periodic  signal \cite{dha} is  as follows. We divide the half cycle of the
signal by  $(2p+1)$  intervals  so  that  each  interval  in  the
horizontal         $\theta$-axis         is        of        size
$(\bigtriangleup\theta/(2p+1))$ with $\omega\bigtriangleup\theta
=\pi$. We define $(2p+1)$numbers $s_{k}$ along the vertical axis
as

\begin{mathletters}
\label{eq.4}
\begin{eqnarray}
s_{k} &=& \frac{[ sin\frac{k\pi}{2p+1} + sin\frac{(k-1)\pi}{2p+1} ]}{2}~
; k = 1,2,...,p
\\s_{p+1} &=& 1
\\s_{p+1+r} &=& s_{p+1-r}~ ; r =1,2,...,p .
\end{eqnarray}
\end{mathletters}
Each      number     $s_k$     is     associated     with     the
interval$\frac{(k-1)\bigtriangleup  \theta}{2p+1}  <\theta   \leq
\frac{k\bigtriangleup\theta}{2p+1}$  with  $k  = 1,2,...,(2p+1)$.
The Eq.(\ref{eq.4}) clearly shows that

\begin{equation}
\label{eq.5}
0<s_{1}<s_{2}<...<s_{p}<s_{p+1}=1>s_{p+2}>s_{p+3}>...>s_{2p+1}>0~.
\end{equation}
Eq.(\ref{eq.5})  states  that in order to reach the maximum value
$(=1)$ of the signal from the zero level we have to have  $(p+1)$
step  up  and  from the maximum to the zero level we have $(p+1)$
step down. This is for the positive half-cycle. For the  negative
half-cycle   similar   constructions  have  been  done  with  the
replacement $s_{k}\rightarrow - s_{k}, \forall k$ and each number
$- s_{k}$ is associated with the  interval  $\bigtriangleup\theta
[1+  \frac{k-1}{2p+1}]<\theta\leq \bigtriangleup \theta [1+ \frac
{k}{2p+1}]$ with $k=1,2,...,(2p+1)$. This  aproximation  for  the
full   one   cycle   of   the  sinusoidal  signal  (as  shown  in
Fig.1) is then repeated for the next successive cycles.

One  may however note that the $\omega$ which we have defined for
this approximated signal is not the same as  that  of  sinusoidal
signal  because  the Fourier transform of sinusoidal signal would
give only one frequency while this  approximated  signal  in  the
Fourier   space   corresponds   to  many  sinusoidal  frequencies
specially because of its sharp discontinuities. Yet we urge  this
approximation  because  in  each  interval  the  equation  become
time-independent.

In  the  future  development  we  associate the index $n$ for the
positive half-cycle and index $m$ for  the  negative.  Index  $i$
will  refer  the  cycle  number. Since the Fokker-Planck equation
Eq.(\ref{eq.3}) in each interval will  be  that  for  a  constant
bias, we can express the conditional probability density function
$p(\xi,\theta   \mid   \xi',\theta')$   in   terms   of  complete
orthonormal set of  eigenfunctions  $u_{n}(\xi)$  satisfying  the
boundary conditions $u_{n}(0) = u_{n}(\Lambda) =0$.

\begin{equation}
\label{eq.6}
p(\xi,\theta\mid\xi',\theta') = \sum_{n} u_{n}^+(\xi) u_{n}^-(\xi')
exp[-\lambda_{n}(\theta-\theta')]~,
\end{equation}
where

\begin{mathletters}
\label{eq.7}
\begin{eqnarray}
u_{n}^\pm(\xi) &=& (2/\Lambda)^\frac{1}{2}exp(\pm s\xi/2)sin\frac{
n\pi\xi}{\Lambda}~  ,\\
\lambda_{n} &=&\frac{n^2\pi^2}{\Lambda^2} +\frac{s^2}{4}~ .
\end{eqnarray}
\end{mathletters}
with $s$ as the corresponding value of $s_{k}$ in the appropriate
interval  where  the conditional probability is being decomposed.
The conditional probability density function in any interval, say
$l$,  can  then  be  calculated  from  the  previous  history  by
convoluting it in each previous intervals:

\begin{equation}
\label{eq.8}
p(\xi_{l},\theta_{l}\mid \xi_{1},\theta_{1}) =
\int...\int d\xi_{l-1}d\xi_{l-2}...d\xi_{2}\prod_{j=2}^{l}
p(\xi_{j},\theta_{j}\mid \xi_{j-1},\theta_{j-1})~.
\end{equation}
For  the  negative  half-cycle  the  calculation  of  probability
density function is similar except that we have  to  replace  the
index  $n$  by  $m$  and  the  probability  density  function  is
decomposed as

\begin{equation}
\label{eq.9}
p(\xi,\theta\mid\xi',\theta') = \sum_{m} u_{m}^-(\xi) u_{m}^+(\xi')
exp[-\lambda_{m}(\theta-\theta')]~,
\end{equation}
where  the expressions for $u_{m}^\pm(\xi)$ and $\lambda_{m}$ are
same as in Eqs.(\ref{eq.7}).

The  survival  probability  at time $\theta$ when the particle is
known to start from $\xi = \xi_0$ at $\theta = 0$ is defined as

\begin{equation}
\label{eq.10}
S(\theta\mid\xi_{0}) =\int_{0}^{\Lambda} d\xi p(\xi,\theta \mid \xi_{0},0)~.
\end{equation}

The  first  passage  time density function (FPTDF) $g(\theta)$ is
defined as

\begin{equation}
\label{eq.11}
g(\theta\mid\xi_{0}) = -\frac{dS(\theta\mid\xi_{0})}{d\theta}~.
\end{equation}
Physically,  $g(\theta)d\theta$  gives  the  probability that the
particle arrives at  any  one  of  the  boundaries  in  the  time
interval  $\theta$  and  $\theta  +  d\theta$.  From this density
function one can calculate various moments:

\begin{equation}
\label{eq.12}
<\theta^{j}> = \int_{0}^\infty d\theta \theta ^{j} g(\theta)~.
\end{equation}
From Eq.(\ref{eq.12}) one can easily calculate mean first passage
time(MFPT)    $<\theta>$    and    the   variance   $\sigma^2   =
<\theta^2>-<\theta>^2$ of the density function $g(\theta)$.

It  is  then  quite straightforward to calculate the FPTDF at any
interval of any cycle. We will write down the final formulae:

\label{eq.13}
\begin{mathletters}
\begin{eqnarray}
g_{+}(\theta\mid\xi_{0})=&& C^+_{n_{(2p+1)(i-1)+1}}
\times \lambda_{n_{(2p+1)(i-1)+1}}
\times exp\{-\lambda_{n_{(2p+1)(i-1)+1}}\times [\theta-2(i-1)\bigtriangleup\theta]\}\nonumber\\
\times F_{i-1}(u^-_{n_{(2p+1)(i-1)+1}})\nonumber\\
&&;2(i-1)\bigtriangleup\theta
<\theta\leq(2(i-1)+\frac{1}{2p+1})\bigtriangleup\theta~,
\end{eqnarray}

\begin{eqnarray}
g_{+}(\theta\mid\xi_{0})=&& C^+_{n_{(2p+1)(i-1)+(k+1)}}
\times \lambda_{n_{(2p+1)(i-1)+(k+1)}}
\times exp\{-\lambda_{n_{(2p+1)(i-1)+(k+1)}}[\theta-2(i-1)\bigtriangleup\theta]\}\times\nonumber\\
&&\prod_{j=1}^{k}\{<u^-_{n_{(2p+1)(i-1)+(j+1)}}\mid u^+_{n_{(2p+1)(i-1)+j}}>\}\nonumber\\
&&\times exp[\frac{\bigtriangleup\theta}{2p+1}(k\lambda_{n_{(2p+1)(i-1)+(k+1)}}
-\sum_{j=0}^{k-1}\lambda_{n_{(2p+1)(i-1)+(j+1)}})]\nonumber\\
&&\times F_{i-1}(u^-_{n_{(2p+1)(i-1)+1}})\nonumber\\
&&;(2(i-1)+\frac{k}{2p+1})\bigtriangleup\theta
<\theta\leq(2(i-1)+\frac{(k+1)}{2p+1})\bigtriangleup\theta\nonumber\\
&&;k=1,2,...,(2p-1)~,
\end{eqnarray}

\begin{eqnarray}
g_{+}(\theta\mid\xi_{0})=&& C^+_{n_{(2p+1)i}}
\times \lambda_{n_{(2p+1)i}}
\times exp\{-\lambda_{n_{(2p+1)i}}[\theta-(2i-1)\bigtriangleup\theta]\}\nonumber\\
&&\times A^+(u^-_{n_{(2p+1)i}}, u^+_{n_{(2p+1)(i-1)+1}})
\times F_{i-1}(u^-_{n_{(2p+1)(i-1)+1}})\nonumber\\
&&;(2(i-1)+\frac{2p}{2p+1})\bigtriangleup\theta
<\theta\leq(2i-1)\bigtriangleup\theta~,
\end{eqnarray}

\begin{eqnarray}
g_{-}(\theta\mid\xi_{0})=&& C^-_{m_{(2p+1)(i-1)+1}}
\times \lambda_{m_{(2p+1)(i-1)+1}}
\times exp\{-\lambda_{m_{(2p+1)(i-1)+1}}[\theta-(2i-1)\bigtriangleup\theta]\}\nonumber\\
&&\times <u^+_{m_{(2p+1)(i-1)+1}}\mid u^+_{n_{(2p+1)i}}>\nonumber\\
&&\times A^+(u^-_{n_{(2p+1)i}}, u^+_{n_{(2p+1)(i-1)+1}})
\times F_{i-1}(u^-_{n_{(2p+1)(i-1)+1}})\nonumber\\
&&;(2i-1)\bigtriangleup\theta
<\theta\leq((2i-1)+\frac{1}{2p+1})\bigtriangleup\theta~,
\end{eqnarray}

\begin{eqnarray}
g_{-}(\theta\mid\xi_{0})=&& C^-_{m_{(2p+1)(i-1)+(k+1)}}
\times \lambda_{m_{(2p+1)(i-1)+(k+1)}}
\times exp\{-\lambda_{m_{(2p+1)(i-1)+(k+1)}}[\theta-(2i-1)\bigtriangleup\theta]\}\times\nonumber\\
&&\prod_{j=1}^{k}\{<u^+_{m_{(2p+1)(i-1)+(j+1)}}\mid u^-_{m_{(2p+1)(i-1)+j}}>\}\nonumber\\
&&\times exp[\frac{\bigtriangleup\theta}{2p+1}(k\lambda_{m_{(2p+1)(i-1)+(k+1)}}
-\sum_{j=0}^{k-1}\lambda_{m_{(2p+1)(i-1)+(j+1)}})]\nonumber\\
&&\times <u^+_{m_{(2p+1)(i-1)+1}}\mid u^+_{n_{(2p+1)i}}>\nonumber\\
&&\times A^+(u^-_{n_{(2p+1)i}}, u^+_{n_{(2p+1)(i-1)+1}})
\times F_{i-1}(u^-_{n_{(2p+1)(i-1)+1}})\nonumber\\
&&;((2i-1)+\frac{k}{2p+1})\bigtriangleup\theta
<\theta\leq((2i-1)+\frac{(k+1)}{2p+1})\bigtriangleup\theta\nonumber\\
&&;k=1,2,...,(2p-1)~,
\end{eqnarray}

\begin{eqnarray}
g_{-}(\theta\mid\xi_{0})=&& C^-_{m_{(2p+1)i}}
\times \lambda_{m_{(2p+1)i}}
\times exp[-\lambda_{m_{(2p+1)i}}(\theta-2i\bigtriangleup\theta)]\nonumber\\
&&\times A^-(u^+_{m_{(2p+1)i}}, u^-_{m_{(2p+1)(i-1)+1}})
\times <u^+_{m_{(2p+1)(i-1)+1}}\mid u^+_{n_{(2p+1)i}}>\nonumber\\
&&\times A^+(u^-_{n_{(2p+1)i}}, u^+_{n_{(2p+1)(i-1)+1}})
\times F_{i-1}(u^-_{n_{(2p+1)(i-1)+1}})\nonumber\\
&&;((2i-1)+\frac{2p}{2p+1})\bigtriangleup\theta
<\theta\leq 2i\bigtriangleup\theta~,
\end{eqnarray}
\end{mathletters}
where

\begin{mathletters}
\begin{eqnarray}
\label{eq.14}
C_{n}^+ =&&\int_{0}^{\Lambda} d\xi u_{n}^+(\xi)~,
\end{eqnarray}

\begin{eqnarray}
C_{m}^- =&&\int_{0}^{\Lambda} d\xi u_{m}^-(\xi)~,
\end{eqnarray}

\begin{eqnarray}
A^+(u^-_{n_{(2p+1)i}}, u^+_{n_{(2p+1)(i-1)+1}})=&&exp[-(\frac{\bigtriangleup\theta}{2p+1})\lambda_{n_{(2p+1)(i-1)+1}}]\nonumber\\
&&\prod_{j=1}^{2p}\{<u^-_{n_{(2p+1)(i-1)+(j+1)}}\mid u^+_{n_{(2p+1)(i-1)+j}}>\nonumber\\
&&\times exp[-(\frac{\bigtriangleup\theta}{2p+1})\lambda_{n_{(2p+1)(i-1)+(j+1)}}]\}~,
\end{eqnarray}

\begin{eqnarray}
A^-(u^+_{m_{(2p+1)i}}, u^-_{m_{(2p+1)(i-1)+1}})=&&exp[-(\frac{\bigtriangleup\theta}{2p+1})\lambda_{m_{(2p+1)(i-1)+1}}]\nonumber\\
&&\prod_{j=1}^{2p}\{<u^+_{m_{(2p+1)(i-1)+(j+1)}}\mid u^-_{m_{(2p+1)(i-1)+j}}>\nonumber\\
&&\times exp[-(\frac{\bigtriangleup\theta}{2p+1})\lambda_{m_{(2p+1)(i-1)+(j+1)}}]\}
\end{eqnarray}
\end{mathletters}
and the functions $F_{i}$ are generated through the recursion relation:

\begin{eqnarray}
\label{eq.15}
F_{i}(u^-_{n_{(2p+1)i+1}})=&&<u^-_{n_{(2p+1)i+1}}\mid u^-_{m_{(2p+1)i}}>
\times A^-(u^+_{m_{(2p+1)i}}, u^-_{m_{(2p+1)(i-1)+1}})\nonumber\\
&&\times <u^+_{m_{(2p+1)(i-1)+1}}\mid u^+_{n_{(2p+1)i}}>
\times A^+(u^-_{n_{(2p+1)i}}, u^+_{n_{(2p+1)(i-1)+1}})\nonumber\\
&&\times F_{i-1}(u^-_{n_{(2p+1)(i-1)+1}})~,
\end{eqnarray}
with   $F_{0}(u^-_{n_{1}})  =  u^-_{n_{1}}(\xi_0)$.  The  angular
bracket in any equation implies dot product of the  corresponding
functions,for e.g.,

\begin{equation}
\label{eq.16}
<u^+\mid u^-> = \int_{0}^{\Lambda} d\xi u^+(\xi)u^-(\xi)~.
\end{equation}
The     cycle     variable     $i$     runs     over     positive
integers;i.e.,$i=1,2,3,...$. The positive and negative symbols of
the FPTDF indicate its values over positive and negative part  of
the   cycles   respectively.  In  all  these  expressions,  viz.,
Eqs.(13)-(\ref{eq.15}), any subscript either $n$ or $m$  or  both
wherever  they  appear more than once the summation over them are
implied. The effect of history is explicit in the expressions for
FPTDF, in particular, through recursive  relation  between  cycle
variable  $i$  through  Eq.(\ref{eq.15})  and through the product
symbols in Eq.(13b), Eq.(13e), Eq.(14c) and Eq.(14d). The Eqs.(13)
show how the non Markov character is systematically  incorporated
through   the  convolution  Eq.(\ref{eq.8})  of  the  probability
density  functions.  Once  the  FPTDF   $g(\theta\mid\xi_0)$   is
obtained  from  these  formluae,  the  various  moments  could be
obtained by employing Eq.(\ref{eq.12}). Evaluation of  FPTDF  and
other  relevant  quantities requires sum of infinite series which
must be truncated in order to obtain a final result.  Convergence
of  FPTDF  is ensured by gradually increasing the number of terms
(i.e.,number of eigenvalues) for the calculation. The process  is
truncated  when  MFPT  does not change up to two decimal point of
accuracy with the change of number of terms. \section{results and
discussions} We have already demonstrated \cite{dha}  that  $p=0$
approximation  or  usual  telegraph  signal  approximation of the
sinusoidal signal does not produce  any  resonance.  In  all  the
following  calculations  we  restrict  to  $p=2$ , although $p=1$
approximation of the signal does produce the resonance.  But  the
$p=2$  signal  approximates better than the $p=1$ signal. Further
we take $\xi_0 =\Lambda /2$ , i.e., the particle  is  assumed  to
start from the mid point of the medium.

\subsection{Synchronization between noise and deterministic signal}
The  FPTDF  $g(\theta  ;\omega)$  for  different  frequencies are
plotted as a function of $\theta$ for a given length $\Lambda=20$
in Fig.2. For very low frequency (for  e.g.,$\omega=0.005$),  the
particle  is  acted  on  by  a  constant  force  $s=s_1$  (in the
multi-step periodic approximation) almost all the time before  it
reaches the boundary. The FPTDF curve shows that we have only one
maximum in the entire $\theta$-regime.

As  the  frequency  slowly  increases, the curve shows that apart
from only one profile appearing left and showing a  maximum  like
before,  a  small peak at large $\theta$ also shows up (for e.g.,
$\omega=0.01$). This is because the particle at  this  frequency,
although  most of the time sees a constant bias $s=s_1$, at later
time it might encounter a increased bias $s=s_2$ for a short time
before it reaches the boundary. As $\int g(\theta)  d\theta  =1$,
therefore  the  area  under  the major profile decreases and this
decrement has been compensated by the small peak at the right end
of  the  profile.  In  other  words,  the  $g(\theta)$  for  this
frequency  consists  of  two  curves  superposed over each other.
Further increase of frequency (for e.g.,$\omega=0.02$) shows  the
similar  behavior  as  with $\omega=0.01$ but here the first peak
area decreases and second peak area increases more than that with
$\omega=0.01$.

What we have been doing is that we follow the $g(\theta)$ profile
as  we  increase the frequency and observe a systematic change in
the constituent profiles. As the frequency increases more,  other
biases $s_k$ with $k>2$ start playing their roles in constructing
the FPTDF profile. For e.g.,$\omega=0.03, 0.04$ , we have another
little  peak  at  extreme  right  starts developing. As frequency
further increases, this  last  peak  starts  developing  further,
while  the  first  peak  which  has  been  dominant  at  very low
frequency starts slowly decreasing.

Our  previous  study  with  this scheme \cite{dha} shows that the
MFPT  has  a  minimum  for  this  length   ($\Lambda   =20$)   at
$\omega=0.1$.  We  have  identified  this  frequency  as resonant
frequency because we have argued that the synchronization between
the signal and noise is maximum at  this  frequency  causing  the
enhancement  of  probability  of reaching the boundary at a short
time. The FPTDF profile at resonance evidently shows  a  dominant
single  peak  at  this  frequency.  More smoother and nice single
dominant peak at resonance is observed for higher length [see for
e.g.,Fig.7]. Slightly below this resonance frequency  $(\omega  =
0.07,  0.08,\Lambda  = 20)$ the FPTDF profile contains two peaks,
the maxima of them are close to each other and the peak which has
been dominant at very low frequency diminishes so much that  only
very  close observation of its nature starting from low frequency
would help identify its  existence.  We  further  note  that  the
constituent  profiles  of  FPTDF  adjust  themselves  in  a  very
distinct way, as the frequency increases, in order to  produce  a
dominant maximum at the resonance.

In  order  to  explain  synchronization  we  must  interpret  the
cooperation in terms of correlation  between  the  components  of
FPTDF  profile.  For  this  purpose we start looking at the FPTDF
profiles for the frequencies higher than the resonant  frequency.
We  have argued that at resonance, the synchronization becomes at
its peak.  After  this  resonant  frequency,  another  small  and
distinct  peak  (for  e.g., $\omega=0.13$) starts developing. The
area of the dominant peak starts reducing. As frequency increases
further beyond the resonance  frequency,  more  number  of  peaks
start   coming   up   over   a   little   background  (for  e.g.,
$\omega=0.2$). After the resonance, the peaks are quite  distinct
and  identified separately over the visible background. Following
the dominant peak beyond resonance, it is natural to identify the
area of the peak as a degree  of  synchronization.  Our  previous
study  with  this scheme \cite{dha} shows that the MFPT increases
beyond the resonance. In these figures we find that the  dominant
peak whose area is maximum at resonance decreases as we go beyond
the   resonant   frequency.   In   other  words,  the  degree  of
synchronization  is  getting   reduced   beyond   the   resonance
frequency.  Clearly  we  can  find  the  areas  of the individual
distinct peaks and the background . The profile of the background
is obtained by fitting the minima of  the  FPTDF  profile.  Going
backwards from high frequency side it is clearly visible that the
background area diminishes as we approach the resonance ; Most of
the  area  is  taken  up  in producing the major dominant peak at
resonance.  The  question  naturally  would  be   how   one   can
distinguish  the  components, namely the background and the peaks
as we decrease the frequency below the resonant frequency.

The clue to the answer of this question lies in the FPTDF profile
at  very  low  frequency. As we have already noticed that at very
low frequency we have only one peak (for  e.g.,  $\omega=0.005$).
The  area  of  this  curve  gradually  decreases as the frequency
increases and approaches its  minimum  at  resonance.  After  the
resonant  frequency,  it  appears clearly as background on top of
which the peaks are occuring. It is evident that this curve  does
not  really play a role in making the dominant peak at resonance.
The other two peaks which appear at larger frequencies  (say,  at
$\omega  = 0.07, 0.08$) start adjusting themselves constructively
to produce a large dominant peak at  resonance  $(\omega  =0.1)$.
Once  we  identify  the background in the low frequency regime by
following the component corresponding to  the  curve  at  $\omega
=0.005$,  the  area  of  the  rest of the profile of FPTDF at any
frequency over the background could be easily calculated.  As  we
have  argued  that  the  two  peaks  really  overlap  each  other
especially   near   the   resonance   and    adjust    themselves
constructively,  the  area  under  these  should  be a measure of
synchronization before the resonance.

Having  identified  the  background,  so  far as the resonance is
concerned, we try to fit this component in each FPTDF profile. As
mentioned before, for very low frequency the full FPTDF gives the
profile of the background. We have also noted that  at  extremely
low  frequency,  in  our  multi-step  periodic signal scheme, the
particle experiences a constant bias $s=s_1$ throught the  entire
time. But no simple analytical formula of FPTDF in such situation
exists.   However,  for  semi-infinite  domain,  if  we  have  an
absorbing boundary at $\xi=\Lambda$ and the particle starts  from
$\xi=\Lambda/2$  the  exact  analytic  expression  for  FPTDF was
evaluated by Schrodinger \cite{sch}. The expression reads as

\begin{equation}
\label{eq.17}
g_s(\theta) = \frac{\Lambda}{4(\pi\theta^3)^{1/2}}
exp\{-(\Lambda/2 - s_1\theta)^2/4\theta\}~.
\end{equation}
We  plot this expression and superpose over actual FPTDF curve at
$\omega=0.01$. The plot is shown  in  Fig.  3.  It  appears  that
although  for low $\theta$ this function gives excellent fit, for
large $\theta$ it  overestimates  the  background.  Therefore  we
modify  this  function in the following way to arrive at a better
fit of the background.

\begin{eqnarray}
\label{eq.18}
g(\theta) =&& g_s(\theta) \hspace{1cm} ;0 < \theta \leq \theta_0 \nonumber\\
=&& g_s(\theta) exp\{-\beta (\theta-\theta_0)^2\} \hspace{1cm} ;\theta >\theta_0
\end{eqnarray}
The  fitted  curve  with  $\theta_0  =  45.14,  \beta= 5.66\times
10^{-4}$ is shown in the same  figure  [Fig.3(b)].  As  frequency
changes  and approaches resonance, $\theta_0$ and $\beta$ change.
The area under this curve, $P_b$, is plotted in Fig.4a and  found
decreasing  before resonance. After resonance, the profile of the
visible background is obtained by joining  the  minima  of  FPTDF
profile  at  each  frequency.  When  the  minima are plotted as a
function   of   $\theta$,   it   fits   with   the   same   curve
Eq.(\ref{eq.18})  with  $\theta_0=0$ and $s_1=0$. Only $\beta$ is
found changing slowly and the area  under  this  curve  is  found
increasing  as the frequency increases beyond resonance (Fig.4a).
This concludes the characterization of the background.

The  area  of  the  peaks over the background, P, responsible for
constructing resonance before the  resonance  frequency  is  also
plotted  in  the  same  figure (Fig.4a) and found increasing with
frequency. At resonance,  as  expected  this  area  denoting  the
degree  of  synchronization attains a peak. As we follow the area
of the dominant peak beyond resonance,  it  is  found  decreasing
with  frequency  showing the degree of synchronization is getting
reduced as we go beyond the  resonant  frequency.  This  is  also
demonstrated in Fig.4a.

Further,  beyond  the  resonance, as mentioned before, apart from
the dominant peak there are other smaller peaks appearing at  the
top  of  the  background.  The areas, $P_1, P_2$, under first two
sucessive peaks beyond the dominant peak are also  plotted  as  a
function  of  frequency  in  the  Fig.4b.  These curves also show
non-monotonous behavior.  This  shows  that  the  synchronization
might  also  occur  at  higher  frequency but as the maximum area
covered by these peaks are much less, they denote synchronization
of   lesser   degree.   This   completes   the   explanation   of
synchronization behavior between noise and deterministic periodic
signal as we change the frequency.

\subsection{Scaling of FPTDF at resonance}
We have already noted that the FPTDF exihibits a dominant peak at
resonance.  This  feature  is  observed  for  any  length. In our
previous study \cite{dha} we have found a simple relation between
the resonant frequency $\omega^*$ and the length  of  the  medium
$\Lambda$. The expression reads as

\begin{equation}
\label{eq.19}
\omega^* = 2/\Lambda ~.
\end{equation}
It  is  further  interesting to observe that the normalised FPTDF
$g(\theta)/\omega^*$ when plotted as a  function  cycle  variable
[$\omega^*  (\Lambda)\theta$],  they overlap over each other. The
curves for different lengths are plotted in  Fig.5.  This  figure
clearly  shows  that  at  resonance cycle variable is the correct
argument to describe the resonance feature. We may  further  note
that  such scaling of the argument of FPTDF would not be possible
for any frequency other than the resonant frequencies because any
frequency which is not the resonant frequency for one length  may
turn  out  to be the resonant frequency for some other length and
the features of FPTDF are different for resonant and off-resonant
frequencies as has been observed from Fig.2.

Having  found  the proper scaling of the argument of FPTDF, it is
natural   to   enquire   whether    the    FPTDF,    $f    \equiv
g(\theta)/\omega^*$  can  also  be  scaled properly, so that once
$f(x)$ with $x=\omega^* \theta$ is  found  for  one  length,  the
function $f(x)$ can be obtained for any arbitrary length.

In  order  to investigate this issue, first we observe that going
from $f(x;\Lambda_1)$ to $f(x;\Lambda_2)$, where $\Lambda_1$  and
$\Lambda_2$  are  two  different  lengths we have to multiply one
function by different  amount  depending  on  the  value  of  the
argument. Thus if at all any scaling relation exists, this should
be of the form

\begin{equation}
\label{eq.20}
f_{\mu\Lambda}(x) = f_{\Lambda}(x) \mu^{\alpha(x)}
\end{equation}

It is indeed found to be true with $\alpha(x)$ given in Fig.6. We
note  that $\alpha(x)$ is universal. From the relation (20) it is
obvious that

\begin{equation}
\label{eq.21}
\Lambda^{-\alpha(x)} f_{\Lambda}(x) = [\mu\Lambda]^{-\alpha(x)} f_{\mu\Lambda}(x).
\end{equation}
The      Eq.(\ref{eq.21})     shows     that     the     function
$\Lambda^{-\alpha(x)} f_{\Lambda}(x)$ is independent of $\Lambda$
and  it  is  again  universal.  We  verify  this  result  in  our
calculation of FPTDF for different lengths and call this function
$U(x)$.  The  $lnU(x)$  is  also  plotted in Fig.6. Thus we could
express $f_{\Lambda}(x)$ at resonance in terms of  two  universal
functions $U(x)$ and $\alpha(x)$ as

\begin{equation}
\label{eq.22}
f_{\Lambda}(x)= U(x) \Lambda^{\alpha(x)}
\end{equation}
The  expression(22)  clearly  isolates the dependence of FPTDF on
length of the medium $\Lambda$ at resonance. Therefore,  for  any
arbitrary  length  $\Lambda$  the  FPTDF at rsonance can be found
from this simple expression.

We next fitted the universal functions $\alpha(x)$ and $lnU(x)$
with the simple functions as
\begin{mathletters}
\begin{eqnarray}
\label{eq.23}
\alpha(x)=-a_{\alpha}\psi(x) - c_{\alpha}\hspace{1cm},\\
lnU(x)= a_U\psi(x)- c_U  \hspace{1cm},
\end{eqnarray}
\end{mathletters}
where
$a_{\alpha},  a_U,  c_{\alpha},  c_U$  are positive constants and
$\psi(x)= (1-\frac{x}{x_m})^2/x$ with $x=x_m$ being the  position
of the maximum of $f(x)$. With these approximate expressions (23)
the  normalised  FPTDF $f_{\Lambda}(x)$ at resonance takes a very
neat expression

\begin{equation}
\label{eq.24}
f_{\Lambda}(x) = C (\frac{\Lambda}{\Lambda_0})^{\alpha(x)} \hspace{1cm},
\end{equation}
with  $\Lambda_0  = 10.312, C=0.5148$. The normalisation of FPTDF
suggests that $\int f_{\Lambda}(x) dx =1$.  It  is  indeed  found
that  this  approximate  form  (24)  would  result  normalisation
constant, for the lengths  we  consider,  almost  independent  of
$\Lambda$   and   very  close  to  unity.  The  approximate  form
$f_{\Lambda}(x)$ is plotted for a typical length $\Lambda=40$ and
compared with actual data  points  in  Fig.7.  It  is  giving  an
excellent  agreement.  Further  with this form, the MFPT is found
for different $\Lambda$.  It  also  yields  linear  behaviour  in
agreement with our previous study \cite{dha}. This shows that not
only  we  have  found a scaling behaviour (20) which takes $f(x)$
from one length to the other, it is also  possible  to  obtain  a
simple  expression  (24)  for  $f_{\Lambda}(x)$ for any arbitrary
length.

\section{concluding remarks}
We consider a diffusive transport process perturbed by a periodic
signal  in continuous one dimensional medium having two absorbing
boundaries. No perturbation of the signal amplitude is assumed in
this formulation. The scheme we have adopted in this  formulation
is  different  from  the  finite  difference  method to solve the
partial  differential  equation  Eq.(\ref{eq.3})  as  adopted  in
\cite{fle}.  The  cooperative behaviour between the deterministic
periodic signal and random noise leading to  coherent  motion  is
demonstrated  in  terms of FPTDF profile. We show explicitly that
synchronization between the signal and random noise is maximum at
resonance where FPTDF exhibits a major dominant peak.

An  important  characteristic  that  we observe is that the cycle
number is the correct argument to describe FPTDF at resonance. We
also show that there exists a scaling relation between  FPTDF  at
various  lengths through some universal function $\alpha(x)$. The
exact expression for FPTDF at resonance is obtained in  terms  of
two  universal functions which clearly isolates its dependence on
the length of  the  medium.  This  is  an  universal  feature  of
coherent  stochastic resonance. Further a nice simple approximate
form for FPTDF at resonance is obtained  which  is  shown  to  be
fairly  accurate  in  explaining  the behavior at resonance. This
form may be of use in obtaining quicker  result  in  cases  where
more complex situation has been called for.

Although  we  restrict  our calculation with a three-step ($p=2$)
periodic signal, the formulation is quite general and  applicable
for  any  approximation  with  an arbitrary number of steps. This
formulation can also  be  applied  to  any  arbitrary  continuous
periodic signal. For $(p=1)$ periodic signal the background could
be  identified  as  before.  The  resonance occurs for the length
$\Lambda=20$ exactly at the  same  frequency  $(\omega=0.1)$.  We
obtain  a  major  dominant peak at resonance. The only difference
that we observe is  that  the  component  of  FPTDF  just  before
resonance  which makes the peak at resonance constitutes only one
peak instead of two as  with  $p=2$.  The  reason  for  which  is
tailored   as   each  half-cycle  of  the  sinusoidal  signal  is
approximated with three steps for  $p=1$  as  against  with  five
steps   with   $p=2$.  However,  after  resonance  frequency  the
behaviour is similar as with $p=2$ signal.

The  peak  positions of normalised FPTDFs at resonance occur very
near to a quarter of a cycle. From  Fig.5  we  observe  a  slight
deviation  of  the  peak  positions  but  we  believe that if the
sinusoidal  signal  is  approximated  by  more  than   three-step
periodic signal, the position of all the peaks will be the same.

\begin{figure}
\caption{Sinusoidal   signal(dashed   curve)   and   approximated
three-step(p=2) periodic signal(solid curve)  for  the  full  one
cycle as a function of $\theta$.}
\label{fig.1}
\end{figure}

\begin{figure}
\caption{(a)  FPTDF  $g(\theta)$  as  a  function of $\theta$ for
$\Lambda=20$  before and on resonance for frequencies $\omega$ =
0.005, 0.01, 0.02, 0.03, 0.04, 0.07, 0.08 and 0.1 (resonant)
(b) FPTDF $g(\theta)$ as a function
of  $\theta$  for  $\Lambda=20$   after   resonance   for
frequencies  $\omega$ = 0.13, 0.15, 0.2 and 0.3
respectively, [$p=2,\xi_0=\Lambda/2]$.}
\label{fig.2}
\end{figure}

\begin{figure}
\caption{  FPTDF  $g(\theta)$  as  a  function  of  $\theta$  for
frequency $\omega =  0.01$  and  $\Lambda  =  20$(filled  circle)
$[p=2,\xi_0=\Lambda/2]$ (a) The function $g_s(\theta)$ in Eq.(17)
as  a function of $\theta$ (dashed curve) (b) The fitted function
$g(\theta)$ in Eq.(18) as a function of $\theta$ (solid curve).}
\label{fig.3}
\end{figure}

\begin{figure}
\caption{(a) The area of the background $P_b$ (filled square) and
the  area  of the dominant peak $P$ (filled circle) as a function
of frequency $\omega$ (b) The areas $P_1, P_2$ (filled circle) as
a function of frequency $\omega$.}
\label{fig.4}
\end{figure}

\begin{figure}
\caption{The  dominant  peaks of $g(\theta)/\omega^*$ at resonant
frequencies for different lengths ($\Lambda =  30,  35,  40,  45,
50$)  as  a  function of $\omega^*\theta$. The lowermost curve is
for $\Lambda = 30$, and upper curves correspond to higher lengths
respectively, $[p=2,\xi_0=\Lambda/2]$.}
\label{fig.5}
\end{figure}

\begin{figure}
\caption{The   universal   functions  $\alpha(x),  lnU(x)$  as  a
function of $x$. }
\label{fig.6}
\end{figure}

\begin{figure}
\caption{(a) $f_{\Lambda}(x)$   as   a   function   of   $x$   for
$\Lambda=40$ at resonance (filled circle)  $[p=2,\xi_0=\Lambda/2]$
(b) $f_{\Lambda}(x)$  according  to the formula (24) as a function
of $x$ for $\Lambda=40$(solid curve).}
\label{fig.7}
\end{figure}

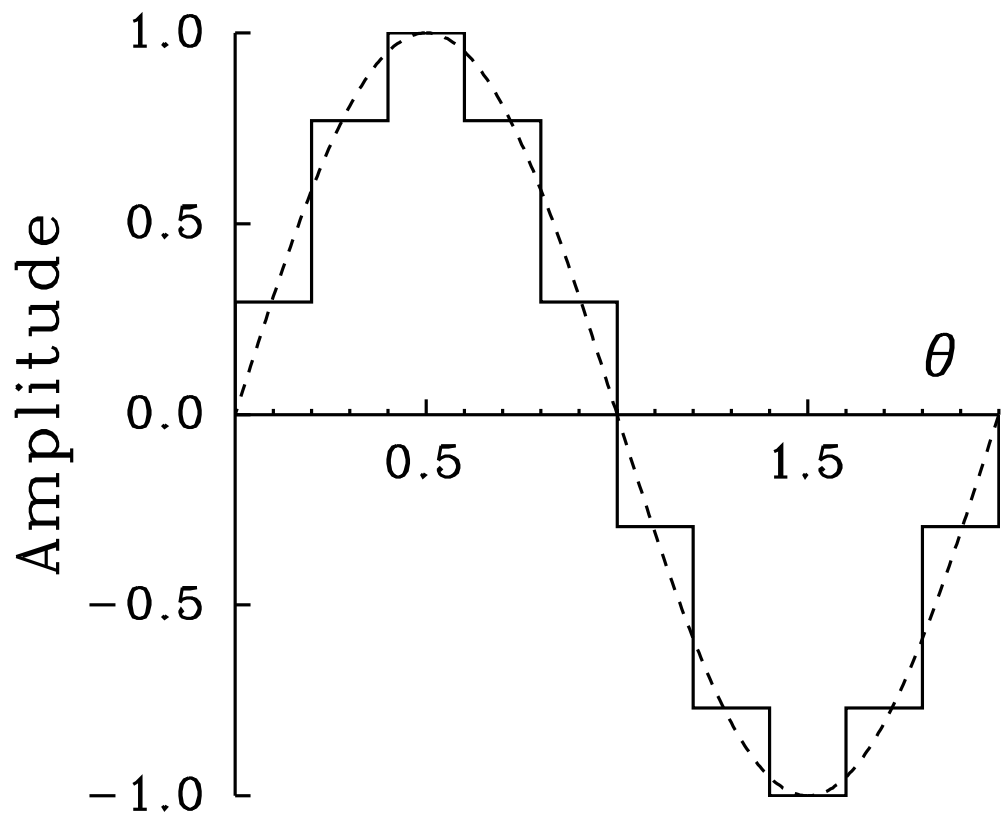
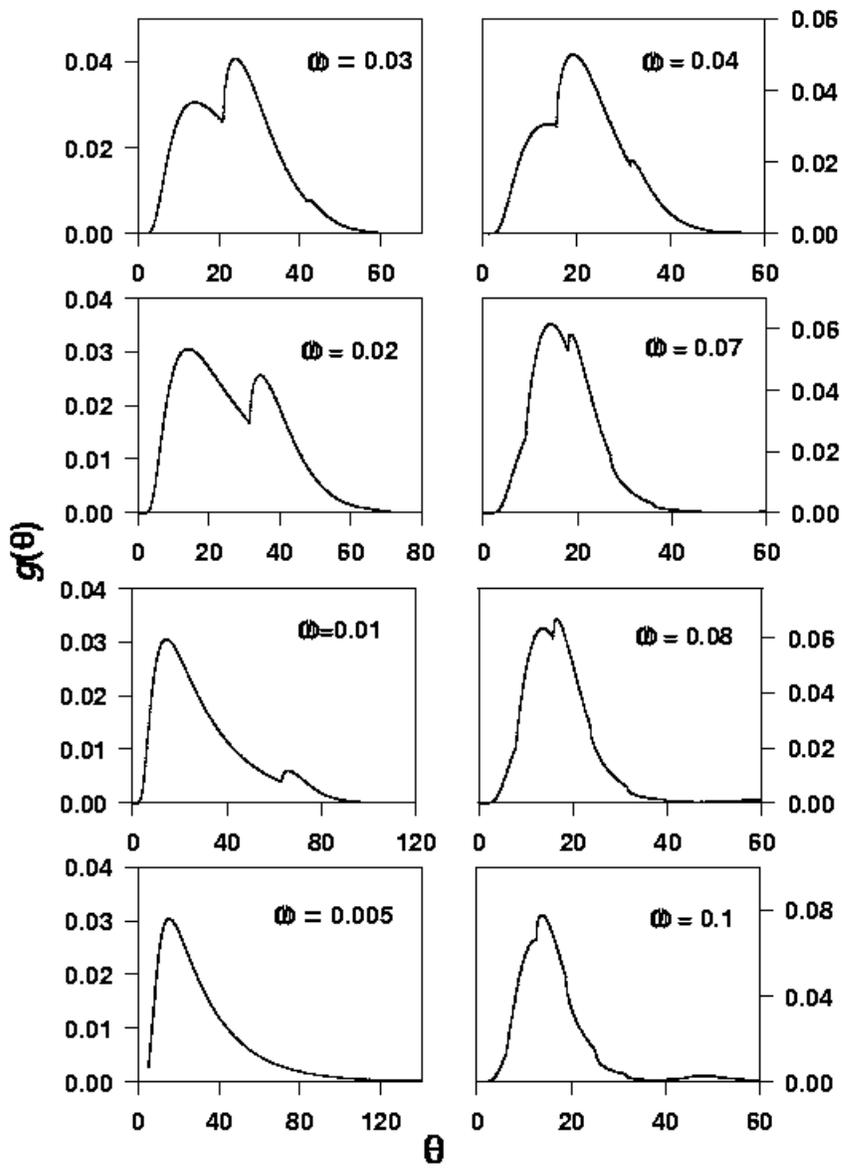
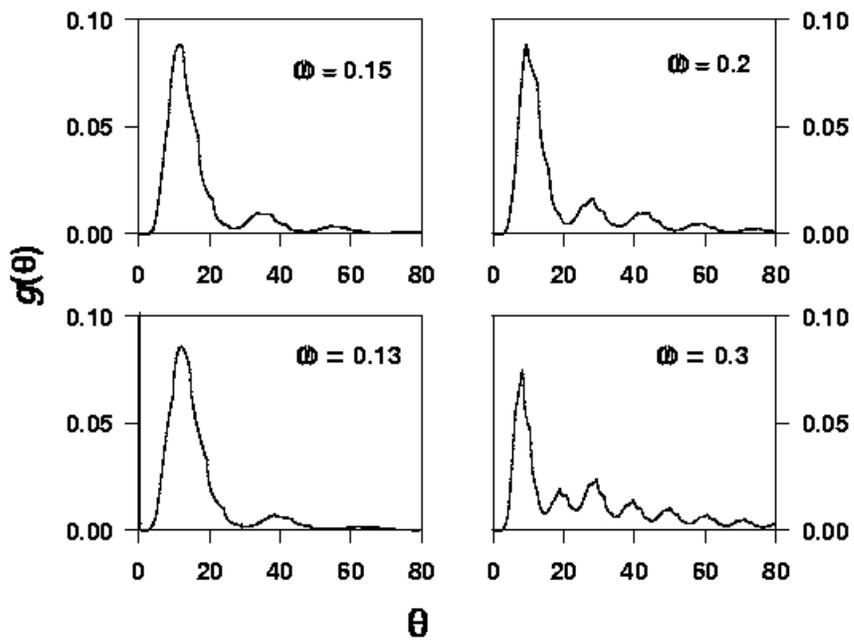
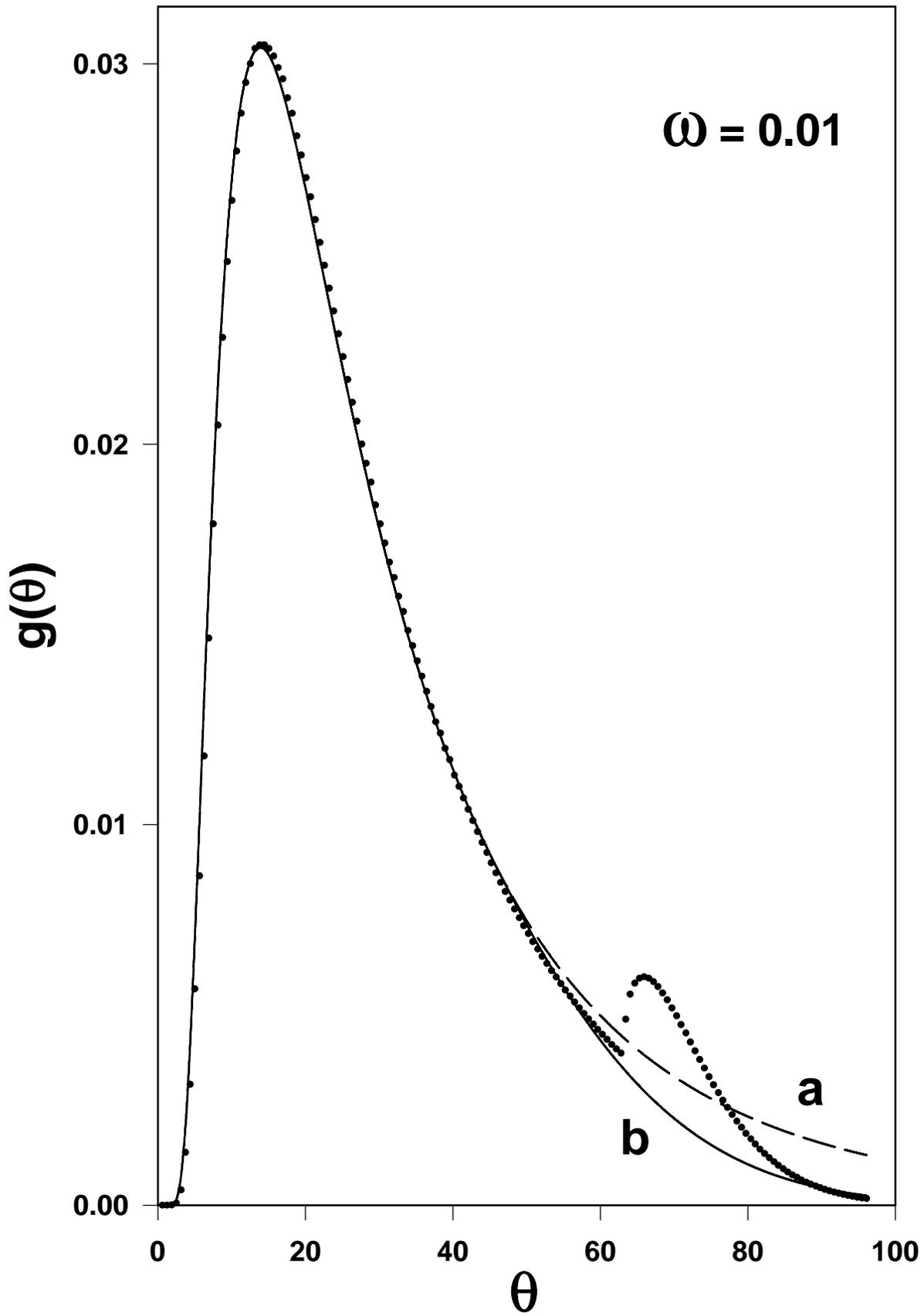
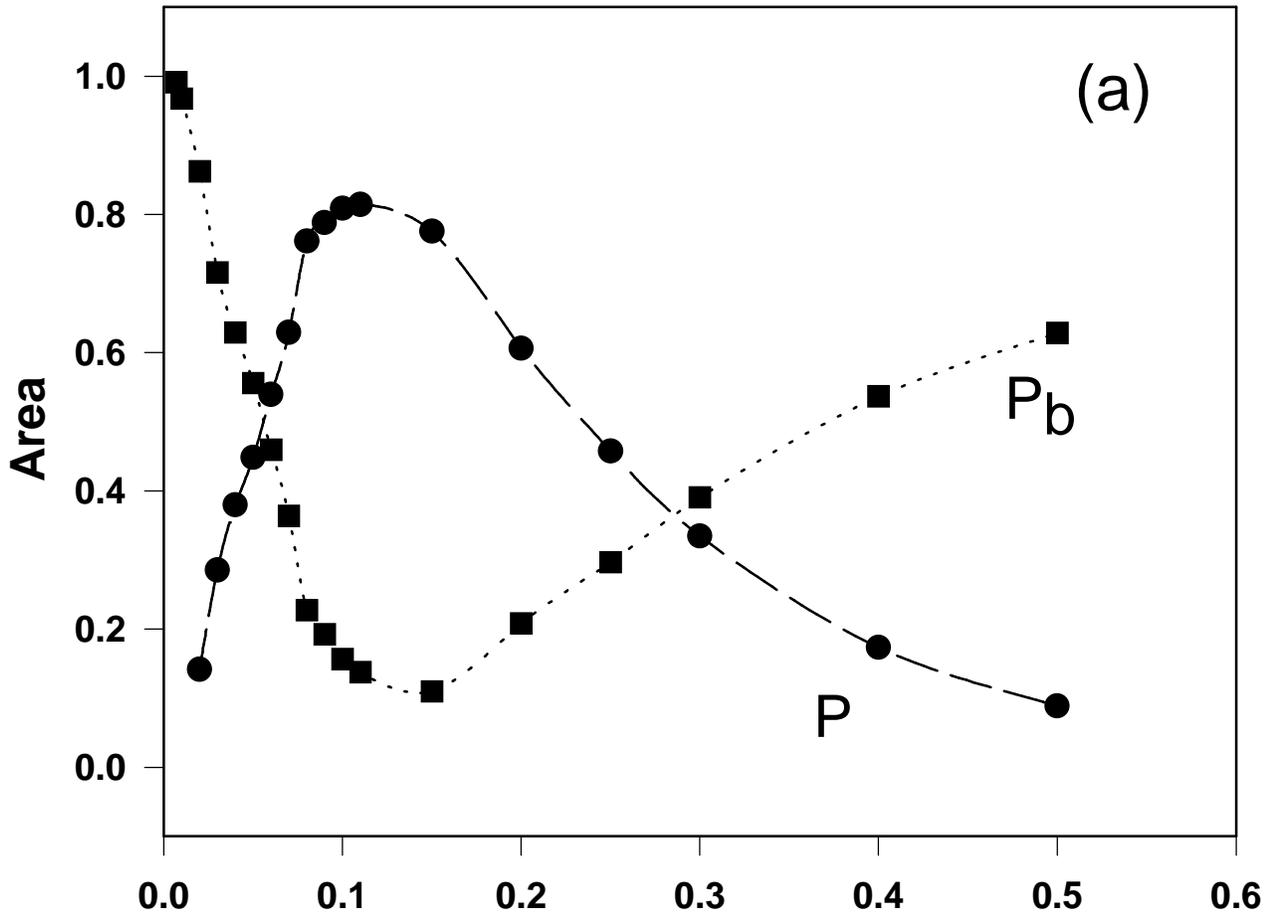
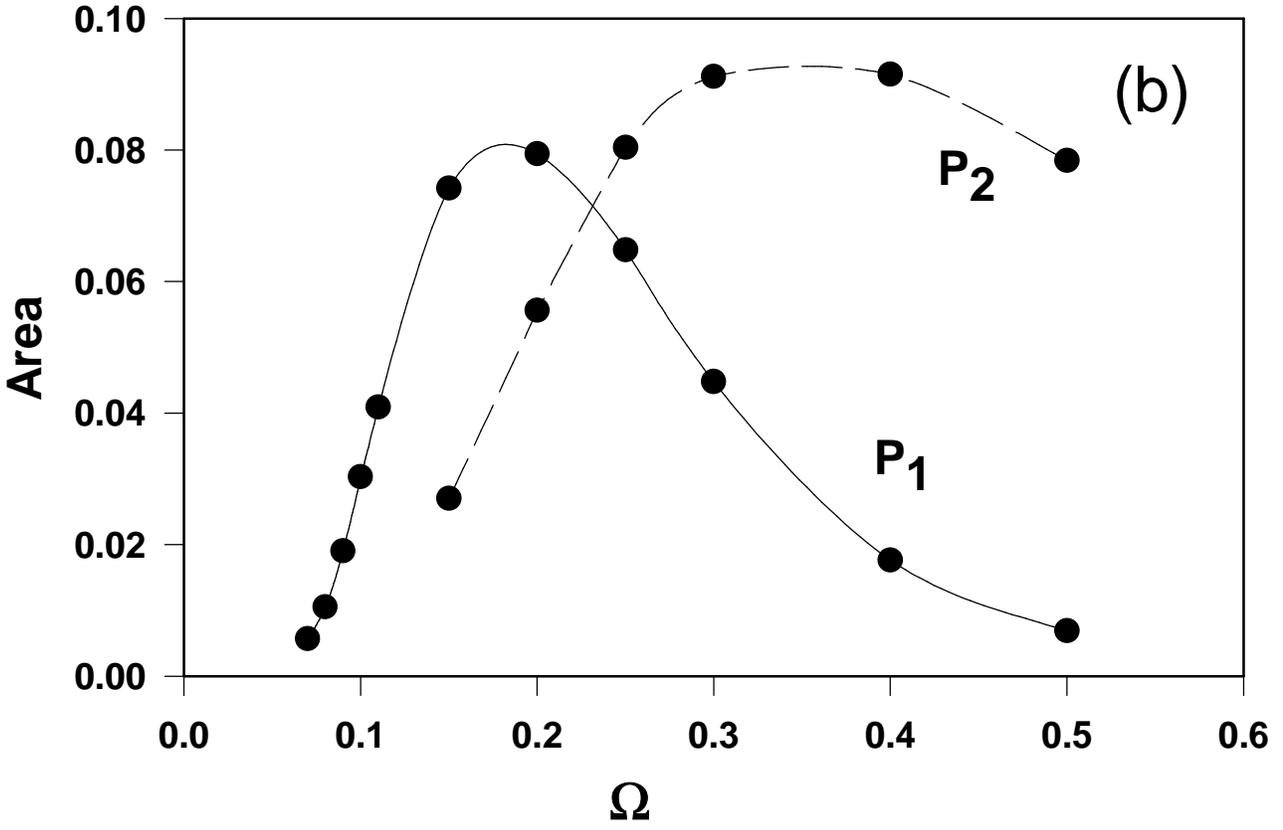
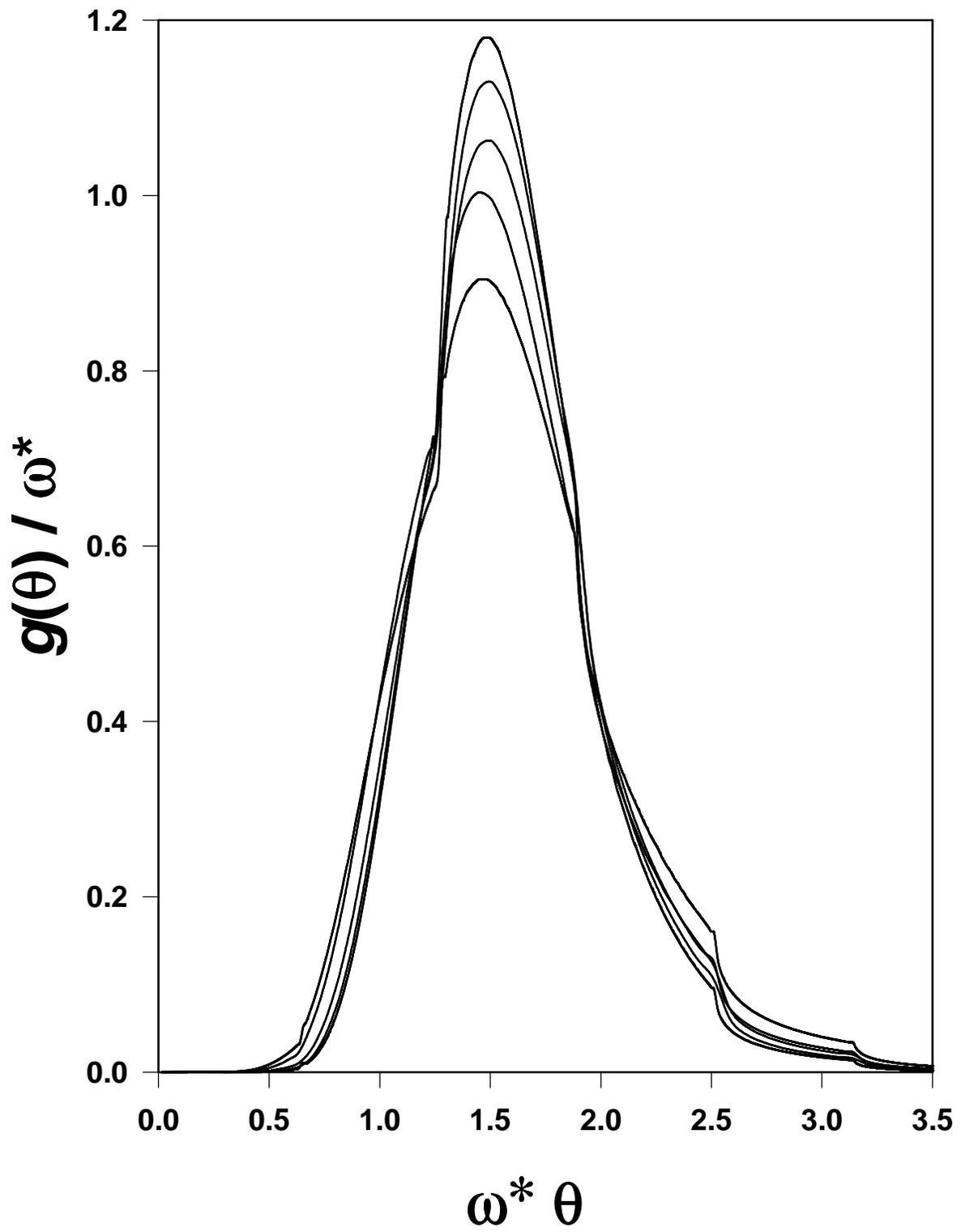
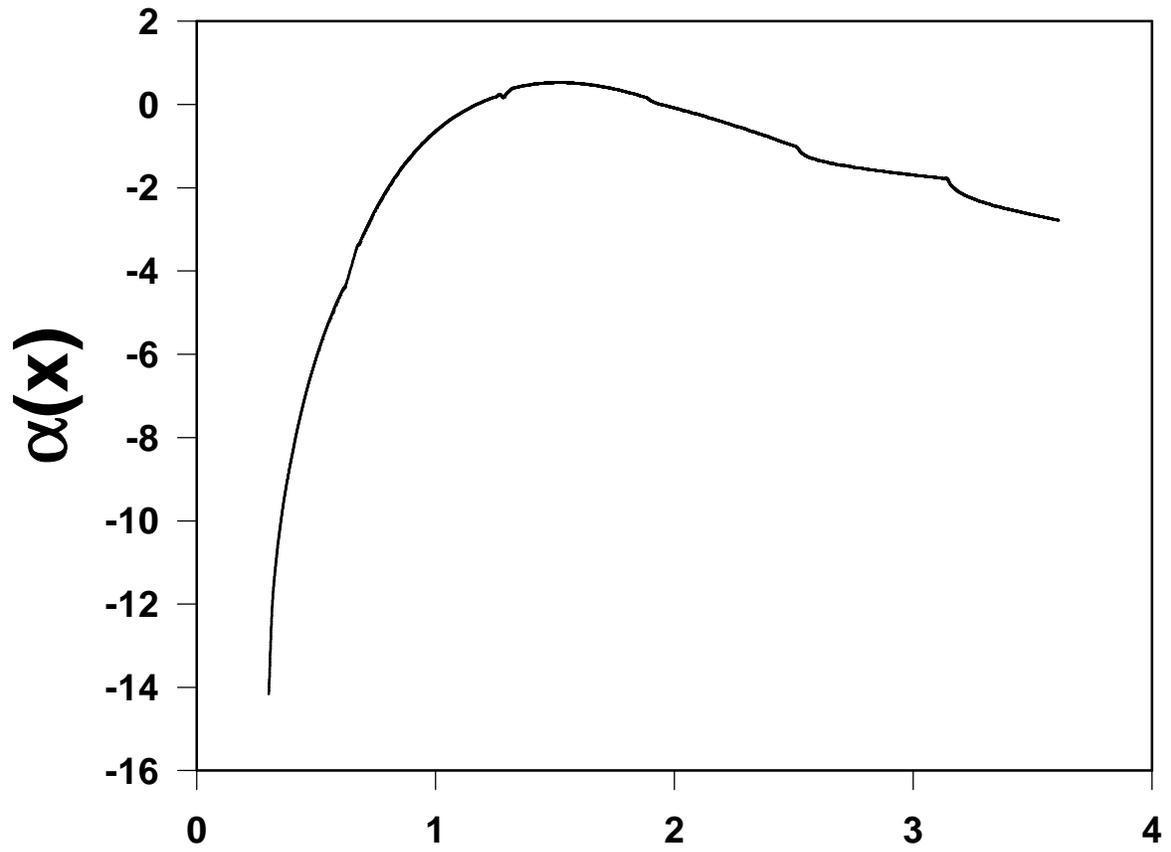
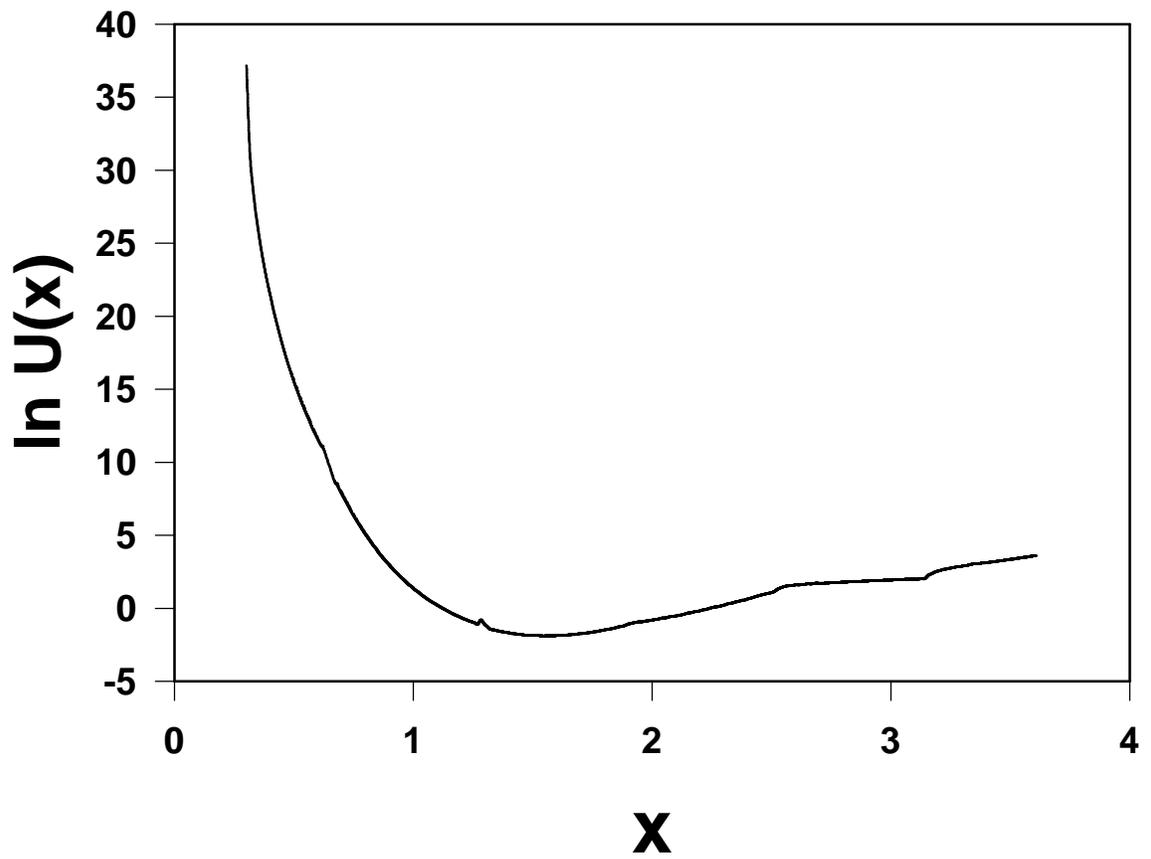
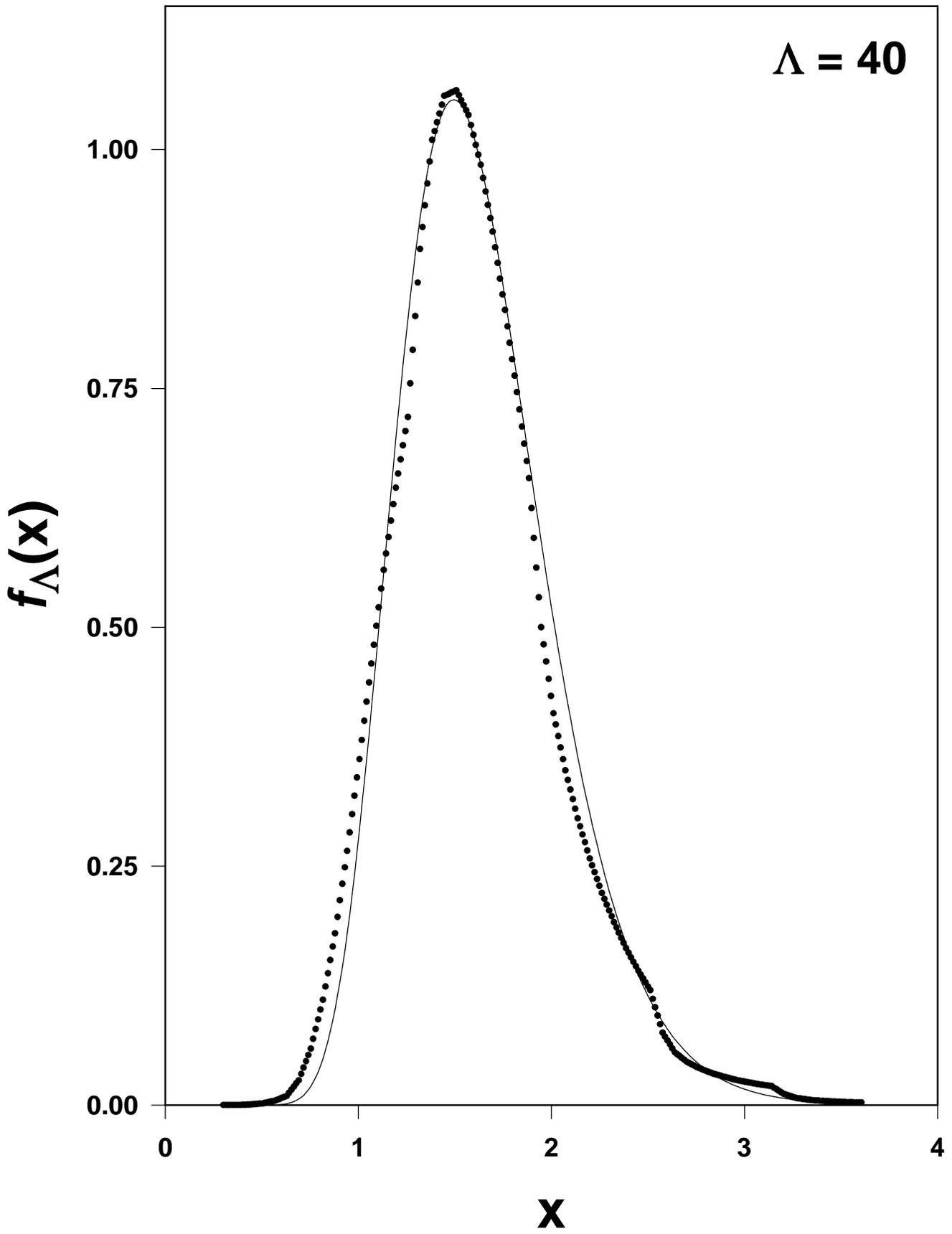
\end{document}